\documentclass[conference,10pt,letterpaper]{IEEEtran}
\usepackage{cite,balance}
\usepackage{amsmath,amssymb,amsfonts}
\usepackage{algorithmic,graphicx}
\usepackage{xcolor,colortbl}
\usepackage{url}
\usepackage[hidelinks]{hyperref}
\hypersetup{pdfauthor={},pdftitle={AscendFHE: Efficient Fully Homomorphic Encryption on Ascend NPUs},pdfsubject={},pdfkeywords={}}
\usepackage{booktabs,tabularx,array,makecell}
\usepackage{breqn,mathtools}
\usepackage{etoolbox,algorithm,enumitem}

\makeatletter
\patchcmd{\@makecaption}{\scshape}{\upshape}{}{
  \PackageError{AscendFHE}{Could not adjust the table caption shape}{Check the IEEEtran caption definition.}
}
\let\ascend@makecaption\@makecaption
\long\def\@makecaption#1#2{%
  \begingroup
  \let\footnotesize\small
  \ascend@makecaption{#1}{#2}%
  \endgroup
}
\makeatother

\begin{document}
\title{AscendFHE: Efficient Fully Homomorphic Encryption on Ascend NPUs}
\author{
\IEEEauthorblockN{Honghui You\textsuperscript{1,2}\qquad
Haoping Yang\textsuperscript{1}\qquad
Wenzhe Wang\textsuperscript{2}\qquad
Lei Ju\textsuperscript{2,1\textdagger}}
\IEEEauthorblockA{\textsuperscript{1}\textit{School of Cyber Science and Technology, Shandong University}, Qingdao, China\\
\textsuperscript{2}\textit{Quan Cheng Laboratory}, Jinan, China\\
yhh21@mail.sdu.edu.cn\quad 202621183@mail.sdu.edu.cn\quad
ts-wangwzh@qcl.edu.cn\quad julei@sdu.edu.cn}
\thanks{\textsuperscript{\textdagger}Lei Ju is the corresponding author.}
}
\maketitle

\begin{abstract}
Fully homomorphic encryption enables cloud servers to run neural networks
on encrypted inputs. In CKKS, inference requires costly arithmetic on
high-degree polynomials. On Ascend neural processing units, preparing
operands and reconstructing intermediate results add to this cost,
while limited local memory constrains parallel execution.
We present AscendFHE, a CKKS backend that coordinates Ascend's matrix
engines and vector cores for bootstrapping and encrypted inference.
It uses balanced-byte encoding and exact modular
reconstruction to reduce vector work and packs intermediate results in
the next operator's input format to avoid redundant transfers. Its
two-task pipeline reuses shared buffers after their last read to
overlap matrix and vector execution within the local memory budget.
On Ascend 910B3, AscendFHE achieves a
4.61$\times$ bootstrapping speedup and 1.59$\times$--2.00$\times$
encrypted-inference speedups across ResNet-20, BERT-base, and Llama 3 8B
over our Vector-based Ascend port of Cheddar.
\end{abstract}

\begin{IEEEkeywords}
AI and Machine Learning, Security \& Privacy.
\end{IEEEkeywords}

\section{Introduction}
\label{sec:introduction}
\begingroup
\setlength{\parskip}{0pt}

Large language models (LLMs) increasingly support applications
in healthcare, finance, and personalized services, where inputs
often contain sensitive user information. When inference is
outsourced to a cloud service, conventional execution requires
the provider to access these inputs in plaintext. This exposure
motivates inference methods that preserve input confidentiality
throughout computation.

Fully homomorphic encryption (FHE) enables computation on encrypted
inputs without decryption~\cite{gentry2009fully}, supporting private
neural network inference~\cite{gilad2016cryptonets}.
The client encrypts inputs before sending them to the server,
which evaluates the model on ciphertexts and returns encrypted
predictions for the client to decrypt. Among FHE schemes,
CKKS~\cite{cheon2017homomorphic} supports approximate arithmetic
on real-valued data, making it suitable for the numerical
computations underlying neural network inference.

However, CKKS incurs substantial computational and memory
overhead because operations on encrypted values require
arithmetic on high-degree polynomials with large coefficient
moduli. TensorFHE~\cite{fan2023tensorfhe} reports that ciphertext--ciphertext
multiplication accounts for roughly one-third of encrypted ResNet-20
inference time. Its kernel-level breakdown attributes
92.1\% of ciphertext multiplication time to the number-theoretic
transform (NTT) and its inverse.
To accelerate NTTs, F1 uses a four-step pipeline in an
ASIC~\cite{samardzic2021f1}, and SAM decomposes large transforms into
smaller NTTs on an FPGA~\cite{wang2023sam}. On GPUs,
WarpDrive~\cite{fan2025warpdrive} and HyperDrive~\cite{fan2026hyperdrive}
map NTT stages to Tensor Cores.

We explore whether neural processing units (NPUs) can similarly use
their matrix engines for NTTs.
Ascend NPUs combine high-throughput matrix engines (Cube)
with programmable vector cores (Vector) and software-managed
local memory~\cite{zhou2025squeezing}. Each NTT output is a weighted sum of the input coefficients modulo q. To compute
these sums with Cube's INT8 matrix operations, Vector encodes the
32-bit input residues as signed bytes. It then reconstructs the sums
from Cube's partial results and reduces them modulo the same prime.
Between NTT rounds, Vector passes coefficients to Cube through global
memory (GM)~\cite{ascend_architecture,ascend_crosscore_sync}.
If Vector writes 32-bit residues to GM, a separate kernel must read
them back and pack them for the next Cube round. Vector can avoid
this pass by packing the coefficients while they are still in its
Unified Buffer (UB) and writing the packed INT8 inputs directly to GM.
The same UB also holds incoming partial sums from
Cube~\cite{ascend_ub_capacity}. As Cube produces the next block,
the space Vector still needs to reconstruct and pack the current block
limits how much of the next block it can stage.
How, then, should exact modular arithmetic, intermediate representations,
and buffer lifetimes be coordinated to exploit Cube--Vector
execution within Ascend's local memory budget?

To this end, we present AscendFHE, a CKKS backend that coordinates the path
from Cube operand preparation to consumer inputs. Building on
byte-decomposed matrix formulations~\cite{fan2023tensorfhe,tong2026leveraging},
AscendFHE reduces Vector support work, produces consumer-ready
intermediates, and schedules Cube--Vector overlap within the UB
budget.
Our contributions are:
\begin{itemize}[leftmargin=*,itemsep=2pt,topsep=3pt,parsep=0pt]
\item \textbf{Exact Cube--Vector arithmetic.}
We develop an exact arithmetic path for Ascend that combines
balanced-byte encoding, two-word reconstruction, and
reconstruction-oriented output layouts. Balanced-byte encoding
eliminates runtime input-sum compensation, achieving summed
kernel-time speedups of 1.22$\times$ for NTT and 1.14$\times$ for INTT.
\item \textbf{Consumer-driven operator composition.}
We directly pack outputs from NTT rounds and basis conversion
(BConv) into inputs for the following NTT and fuse Rescale
post-processing into NTT finalization. Across all tested paths,
these mechanisms eliminate 23.5\,MiB of logical GM traffic per
path and achieve up to 1.15$\times$ speedup relative to separately executed kernels.
\item \textbf{Cube--Vector pipelining.}
We separate task concurrency from the number of active UB input buffers
and distinguish shared-read completion from output completion.
Combining a two-task limit with shared-read acknowledgments
yields full-interface speedups of 1.15$\times$ for NTT
and 1.12$\times$ for INTT with one active input buffer per AIV.
\item \textbf{End-to-end FHE on Ascend.}
We implement CKKS bootstrapping and encrypted inference for
ResNet-20, BERT-base, and Llama 3 8B on Ascend 910B3.
Relative to our Vector-based Ascend port of Cheddar~\cite{choi2026cheddar},
the integrated backend achieves a 4.61$\times$ bootstrapping
speedup and 1.59$\times$--2.00$\times$ speedups for encrypted inference.
\end{itemize}
\par
\endgroup

\section{Background}
\label{sec:background}

\subsection{Ascend NPU Architecture}
\label{sec:background-ascend}
\begin{figure}[!t]
\centering
\includegraphics[width=\columnwidth,trim=0 0 0 1bp,clip]{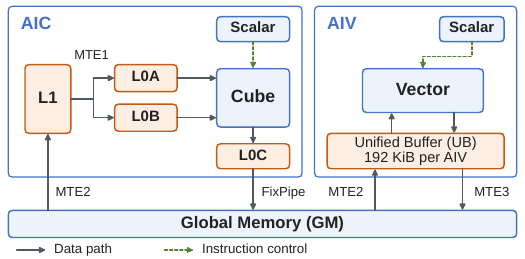}
\caption{Ascend 910B3 compute and memory architecture.}
\label{fig:ascend-architecture}
\end{figure}

Ascend NPUs provide specialized matrix (Cube), vector
(Vector), and scalar execution resources. On our target
Ascend 910B3 (Figure~\ref{fig:ascend-architecture}), Cube and Vector reside on separate AI Cube
(AIC) and AI Vector (AIV) cores. Each core has its own
scalar unit and independent instruction stream~\cite{ascend_architecture}.
Cube accelerates dense matrix multiplication, including
INT8 products with INT32 accumulation. Vector supports
elementwise arithmetic, conversions, and reductions.

The two core types also use separate local storage.
Cube uses dedicated L1 and L0 buffers, while each AIV
provides a 192\,KiB software-managed Unified Buffer
(UB)~\cite{ascend_ub_capacity}. Input staging, output buffering,
and arithmetic temporaries share this UB capacity.
Data exchange between AIC and AIV passes through global
memory (GM), with explicit software synchronization
between producers and consumers~\cite{ascend_crosscore_sync}.
Asynchronous memory transfers can overlap with computation
when dependencies permit; software must coordinate buffer
lifetimes and synchronization to exploit this overlap.

\subsection{CKKS Operators and Kernel Composition}
\label{sec:background-ckks}
CKKS supports addition, multiplication, rotation, and rescaling on
encrypted vectors~\cite{cheon2017homomorphic}. In a residue number
system (RNS), coefficients are stored as remainders modulo small
primes~\cite{cheon2018full}. Basis conversion (BConv) changes the set
of moduli, often between an inverse NTT (INTT) and a forward
NTT~\cite{han2020better}. It uses the least nonnegative (canonical)
or nearest-zero (centered) representative of the source residues.

Key switching combines transforms, BConv, and inner products with
evaluation keys for relinearization and rotation. Rotations also
require automorphisms. Rescaling divides coefficients by the removed
moduli and rounds. In evaluation form, it uses transforms,
RNS basis reduction, and modular corrections~\cite{cheon2018full}.
Bootstrapping restores levels for further
computation~\cite{han2020better}.

Kernel composition exploits these sequences to reduce
intermediate transfers and kernel-launch overhead.
Sequential fusion combines producer--consumer steps, such as NTT
finalization and rescaling corrections. Parallel fusion groups
independent operations sharing inputs, such as products with multiple evaluation
keys, to reuse data~\cite{choi2026cheddar}.

\subsection{Number-Theoretic Transform (NTT)}
\label{sec:background-ntt}
NTT/INTT convert polynomials between coefficient and evaluation
representations modulo a suitable prime, enabling polynomial
multiplication through pointwise modular products.
The inverse transform converts these products back to polynomial coefficients.
Two common approaches are butterfly algorithms and four-step decompositions.
Butterfly
algorithms use $O(N\log N)$ modular operations for a length-$N$ transform.
The four-step algorithm factors the transform length as
\mbox{$N=N_1N_2$} and applies length-$N_1$ and length-$N_2$
transforms along the two dimensions of an $N_1\times N_2$
coefficient matrix, with twiddle-factor multiplication and layout reordering.
Each short transform can be expressed as a modular matrix
product, enabling batched GEMMs that reuse transform
weights across multiple input blocks in the same transform
batch~\cite{fan2023tensorfhe}.

\section{Motivation}\label{sec:motivation}

\noindent\textbf{Exact arithmetic requires architectural adaptation.}
Existing FHE implementations optimize arithmetic for their target hardware.
Cheddar uses widening and high-half multiplication for modular
arithmetic~\cite{choi2026cheddar}. TensorFHE maps byte-decomposed
operands to GPU matrix engines~\cite{fan2023tensorfhe},
while CROSS reduces supporting arithmetic and reordering on
TPUs~\cite{tong2026leveraging}.
Directly porting the obove schemes face two Ascend-specific constraints.
First, on 910B3, Ascend C does not support multiplying two
32-bit integers into a full 64-bit result~\cite{ascend_mull_support}.
Second, peak INT32 Vector addition is 4.61\,TOPS,
compared with 590\,TOPS for INT8 Cube, making encoding,
reconstruction, and modular reduction costly.
These constraints motivate
mapping transforms to Cube while simplifying Vector-side encoding,
reconstruction, and reduction.

\noindent\textbf{Operator boundaries introduce representation costs.}
TPUs coordinate matrix and vector execution within a TensorCore
through compiler-managed asynchronous execution~\cite{tpu_pipelining}.
Ascend instead uses separate AIC/AIV instruction
streams~\cite{ascend_architecture}, with GM-mediated data exchange
and explicit cross-core synchronization~\cite{ascend_crosscore_sync}.
Cube's INT32 partial sums must reach Vector for reconstruction,
while the next Cube round consumes packed INT8 inputs.
Writing the reconstructed coefficients to GM as uint32 values
forces a later kernel to reload and encode them for Cube.
Faster matrix kernels do not remove this extra pass.
This motivates composing operators around their consumers' input
formats to eliminate intermediate uint32 writes and reads.

\begin{figure}[!t]
\centering
\includegraphics[width=0.95\columnwidth]{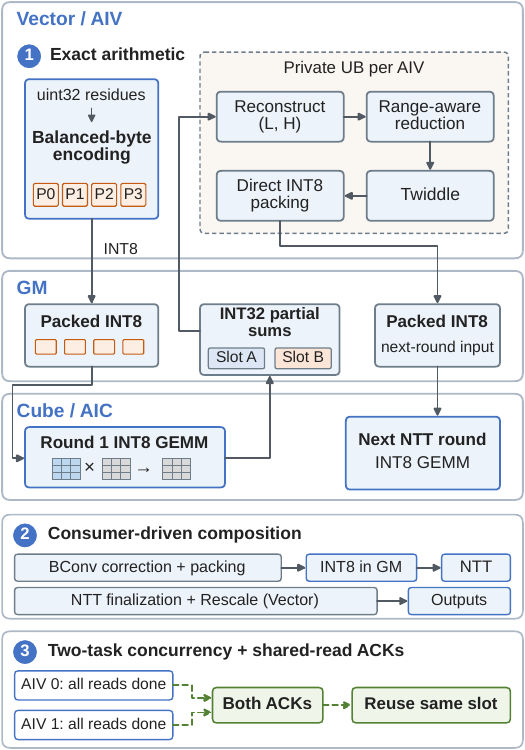}
\caption{Overview of AscendFHE.}
\label{fig:design-overview}
\end{figure}

\noindent\textbf{Limited local storage constrains Cube--Vector overlap.}
Processing one block at a time leaves Cube idle while Vector
reconstructs its output. Overlapping successive blocks can reduce
this waiting, but their intermediate data must coexist with
arithmetic temporaries. Each AIV has only 192\,KiB of
UB~\cite{ascend_ub_capacity}, shared by input staging, reconstruction,
and output buffering. Adding input buffers to overlap more blocks
reduces the space available for reconstruction and output staging.
Operator fusion removes intermediate data transfers, but overlapping
blocks still need storage for their live inputs and arithmetic
temporaries. The achievable overlap therefore depends on when
these data can be discarded and their storage reused.
This motivates coordinating task scheduling and buffer lifetimes
within the UB budget.

\section{Design}
\label{sec:design}

\subsection{Overview}
\label{sec:design-framework}
\label{sec:design-reformulation}

AscendFHE optimizes CKKS bootstrapping and encrypted model evaluation
by coordinating arithmetic, intermediate representations, and scheduling
across Cube and Vector (Figure~\ref{fig:design-overview}).
Exact arithmetic (Section~\ref{sec:design-arithmetic}) reduces Vector
support work through balanced-byte encoding, two-word reconstruction,
and range-aware reduction while preserving exact modular results.
Consumer-driven operator composition
(Section~\ref{sec:design-representation}) arranges Cube's partial sums
for contiguous Vector reads and directly packs UB-resident coefficients
into the next Cube round's INT8 input layout.
Cube--Vector pipelining (Section~\ref{sec:design-execution}) coordinates
these GM-mediated handoffs within the UB budget through segmented
processing and storage reuse.

\subsection{Reducing Vector-Side Overhead}
\label{sec:design-arithmetic}

Following TensorFHE~\cite{fan2023tensorfhe}, we map the two rounds of
short transforms in a four-step NTT to INT8 GEMMs on Cube. Vector packs
the operands for these GEMMs and reconstructs exact modular results from
the INT32 partial sums they produce. Between rounds, Vector applies
twiddle factors to the reconstructed results before packing the next
round's inputs. We reduce this Vector-side overhead through
\textbf{balanced-byte encoding} and \textbf{exact modular reconstruction}.

\subsubsection{Balanced-Byte Encoding}
\label{sec:design-balanced}

Using Cube for modular transforms requires decomposing 32-bit
residues into byte-sized operands. Conventional unsigned-byte
decomposition~\cite{fan2025warpdrive} produces digits outside
Cube's signed INT8 range, requiring a runtime input-sum correction.
We instead encode each centered residue directly as four signed
INT8 digits, preserving its integer value and eliminating this
correction (Algorithm~\ref{alg:balanced-byte}).

\begin{algorithm}[!t]
\caption{Balanced-byte encoding}
\label{alg:balanced-byte}
\begin{algorithmic}[1]
\REQUIRE Residue $x\in[0,q)$; odd modulus $1<q<2^{31}$
\ENSURE Signed INT8 digits $(d_0,d_1,d_2,d_3)$
\STATE $\bar{x}\gets x$; \textbf{if} $x>\lfloor q/2\rfloor$ \textbf{then} $\bar{x}\gets x-q$ \label{line:balanced-center}
\STATE $u\gets\bar{x}+\mathtt{0x80808080}$ \label{line:balanced-bias}
\FOR{$j=0,1,2,3$}
    \STATE $d_j\gets\bigl((u\gg(8j))\mathbin{\&}\mathtt{0xFF}\bigr)-128$ \label{line:balanced-digit}
\ENDFOR
\RETURN $(d_0,d_1,d_2,d_3)$
\end{algorithmic}
\end{algorithm}

The algorithm first centers each residue, then adds the bias
to the full word so that carries are resolved before byte
extraction. The modulus bound keeps the biased value within
the unsigned 32-bit range. The resulting signed digits reconstruct
the centered integer exactly.
We use this representation for NTT inputs and apply it offline
to BAT's folded weights~\cite{tong2026leveraging}, preserving the
integer values of both Cube operands.

\subsubsection{Reconstructing Results with Two 32-Bit Words}
\label{sec:design-reconstruction}

Cube's partial sums fit in INT32, but their weighted sum $S$
may exceed this range. Adjacent columns represent byte positions and
differ in weight by a factor of $2^8=256$. We add a bias $B$, a multiple
of $q$, to make the lower columns nonnegative. Vector keeps the low eight
bits of each column, extracts the carry with an unsigned right shift, and
adds it to the next column. The resulting low and high 32-bit words $L,H$
satisfy \mbox{$S+B=L+2^{32}H$}. Reducing this word pair modulo $q$
recovers $S\bmod q$.

\subsubsection{Simplifying High-Half Multiplication}
\label{sec:design-reduction}

Barrett reduction~\cite{barrett1986implementing} and Shoup constant
multiplication~\cite{harvey2014faster} use the upper half of an
integer product to estimate the quotient. However, Ascend C's Mull API is
unsupported on Atlas A2 products~\cite{ascend_mull_support}.
Following CROSS~\cite{tong2026leveraging}, we split each 32-bit operand
into two 16-bit halves and combine their partial products with
explicit carries to recover the upper 32 bits of the full product.

We specialize this computation using operand bounds: a partial
product is omitted only when one of its operand halves is guaranteed
to be zero, and carry steps are combined only when their intermediate
sum fits in a 32-bit word. These simplifications preserve the
high-half product exactly, leaving quotient estimates and modular
corrections unchanged.

\subsection{Consumer-Driven Operator Composition}
\label{sec:design-representation}

Between NTT rounds, data follows in a Cube--Vector--Cube
path. Cube produces partial sums, Vector reconstructs
coefficients and applies twiddle factors, and the next
Cube round consumes packed INT8 inputs. Each handoff
therefore needs a representation suited to its consumer.
We arrange Cube outputs for Vector reconstruction,
then pack Vector outputs for the next Cube round
while the coefficients remain in UB. We extend this
approach to BConv and Rescale by combining data
preparation with their existing Vector computations.

\subsubsection{Arranging Cube Outputs for Vector}
\label{sec:design-layout}

At the Cube-to-Vector handoff, reconstruction requires
partial sums grouped by byte position. An interleaved
Cube output layout requires Vector to gather these
groups before reconstruction.
We reorder the precomputed weight columns offline so
that Cube writes each partial-sum plane contiguously.
Vector can then read the planes directly for reconstruction.

\subsubsection{Packing Vector Outputs for the Next Cube Round}
\label{sec:design-raw}

The Vector-to-Cube handoff follows reconstruction
and twiddle-factor multiplication. At this point, the
coefficients reside in UB, but the next Cube round
reads packed INT8 inputs from GM. Writing these coefficients
to GM as uint32 values would require reloading and
encoding them in a separate pass. We encode and pack them while they
are still in UB, then write the packed INT8 array to
GM for the next round. This removes the uint32 write
and read and the separate encoding pass.
The second-round kernel starts after the first round
finishes producing the packed array.

Before normalization, the twiddle result $t$ lies in $[0,2q)$.
The standard path reduces $t$ to $[0,q)$ and centers it before encoding
(Algorithm~\ref{alg:balanced-byte}). The optional path skips both steps
by directly encoding $t-q\in[-q,q)$ when it fits in four balanced-byte
digits. Since $t-q\equiv t\pmod q$, the next NTT produces the same
modular result.

\subsubsection{Extending Operator Handoffs to BConv and Rescale}
\label{sec:design-bconv}
\label{sec:design-rescale}

During setup, we check that the separate and joint BConv--NTT paths
produce identical outputs for each eligible shape. We time both paths
end to end and select the joint path when its latency gain meets the
threshold. The joint path maps BConv to
Cube, applies the source-basis correction for the chosen representative,
and packs the target residues directly as INT8 inputs for NTT. This
removes the intermediate uint32 write and read and the separate encoding pass.

For Rescale, let $D$ be the product of the dropped moduli and $r$ the
coefficient-wise centered remainder modulo $D$. We apply INTT to the
dropped limbs and use centered BConv to obtain $r\bmod q$ for each
retained modulus $q$. The output in NTT form is given by
\[
\widehat y_q \equiv \bigl(\widehat x_q -
\operatorname{NTT}_q(r\bmod q)\bigr) D^{-1} \pmod q,
\]
where $\widehat x_q$ is the original NTT-domain input. In finalization,
NTT subtracts the transformed remainder, multiplies by the inverse of
$D$ modulo $q$, and writes the Rescale output.

\subsection{Cube--Vector Pipelining}
\label{sec:design-execution}

We organize execution into tasks,
each producing one Cube output block partitioned into
segments for two AIVs. We allocate two alternating GM
buffers, termed slots A and B, to hold these blocks.

\begin{figure}[!t]
\centering
\includegraphics[width=0.95\columnwidth,trim=0 2bp 0 4bp,clip]{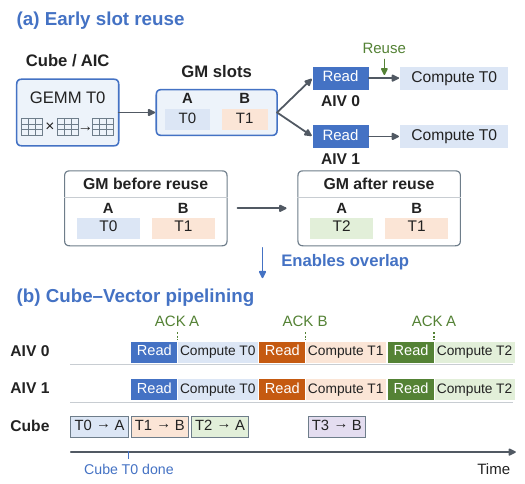}
\caption{Cube--Vector pipelining in intermediate NTTs. $T_i$ denotes
Cube task $i$, and A and B are GM slots for task outputs.
(a) Read ACKs release slots for reuse. Each slot shows the task whose
output it holds. (b) Cube and Vector work overlap. Arrows on the Cube
timeline point from tasks to their destination slots.}
\label{fig:cube-vector-slot-release}
\end{figure}

\paragraph{UB capacity}
\label{sec:design-storage}
\label{sec:design-capacity}
Our evaluated NTT pipeline reserves 185.5\,KiB of each AIV's
192\,KiB UB for input buffers, arithmetic workspace, and output
storage, leaving 6.5\,KiB.
The reservation comprises 112\,KiB for two 56\,KiB input buffers,
65.5\,KiB for workspace, and 8\,KiB for output.
We process Cube outputs in segments to
limit input storage, and reuse temporary storage
between arithmetic and packing when their lifetimes
do not overlap. This allows input staging and
computation to coexist within the UB budget.

Cube can fill the other GM slot while the AIVs process the current block.
Each AIV loads its segment of the next block from GM once its UB input
buffer is free. This allows two tasks in flight with one active UB input
buffer per AIV.

\paragraph{Early slot reuse}
\label{sec:design-acknowledgments}
We implement acknowledgments (ACKs) using cross-core synchronization
primitives~\cite{ascend_crosscore_sync}. Both AIVs acknowledge
when Cube can overwrite a GM slot.
For nonfinal intermediate-NTT tasks, each AIV sends
its ACK after completing all assigned shared reads;
its remaining reconstruction, packing, and output
writes use private UB data.
Once both read ACKs arrive, the slot can be released
without waiting for these remaining operations
(Figure~\ref{fig:cube-vector-slot-release}(a)).
\label{sec:design-terminal}
For the final task, each AIV acknowledges after completing its output
writes. The next NTT round starts after all packed inputs have been written.

\paragraph{Task pipelining}
Early slot release determines when storage is reusable;
the task limit determines whether Cube can submit
another task. This limit caps the number of submitted
Cube tasks not yet acknowledged by both AIVs.
With a limit of two, Cube can fill B for $T_1$ while
both AIVs process $T_0$ from A. To execute $T_2$ next,
Cube must finish $T_1$ and reclaim slot A, which holds
$T_0$'s output. The two read ACKs confirm that neither
AIV needs to read A again. Cube can then write $T_2$'s
output into A while the AIVs finish $T_0$ using their
private UB data.
Together, the two-task limit and early slot release
enable the overlap shown in
Figure~\ref{fig:cube-vector-slot-release}(b).

\section{Evaluation}
\label{sec:evaluation}

\definecolor{EvalTableHeader}{HTML}{EBF1F8}
\definecolor{EvalTableRule}{HTML}{627387}
\newcommand{\evaltablestyle}{%
    \small
    \setlength{\tabcolsep}{4pt}%
    \renewcommand{\arraystretch}{1.12}%
    \arrayrulecolor{EvalTableRule}%
}
\newcommand{\evalnumhead}[1]{\multicolumn{1}{r}{\textbf{#1}}}

\subsection{Experimental Setup}
\label{sec:eval-methodology}

\paragraph{Hardware setup}
We use an Ascend 910B3 and an NVIDIA H100 PCIe 80GB, each on an
Intel Xeon Gold 6530 host. Table~\ref{tab:device-compute-peaks} compares
matrix and integer throughput. The 910B3 uses Cube for GEMM and Vector
for INT32 addition. The H100 uses Tensor Cores and CUDA Cores, respectively.

\begin{table}[!htbp]
\centering
\caption{Theoretical peak throughput derived from CANN for Ascend~\cite{ascend_matmul_tiling,ascend_vector_tiling}
and published by NVIDIA for H100~\cite{nvidia_h100_whitepaper,nvidia_cuda_arithmetic}.}
\label{tab:device-compute-peaks}
\evaltablestyle
\begin{tabularx}{\columnwidth}{>{\raggedright\arraybackslash}Xrrr}
\toprule
\rowcolor{EvalTableHeader}
\textbf{Accelerator}
& \makecell[r]{\textbf{INT8 GEMM}\\TOPS}
& \makecell[r]{\textbf{FP16 GEMM}\\TFLOPS}
& \makecell[r]{\textbf{INT32 add}\\TOPS} \\
\midrule
Ascend 910B3 & 590 & 295 & 4.61 \\
H100 PCIe 80GB & 1513 & 756 & 25.60 \\
\bottomrule
\end{tabularx}
\end{table}

\paragraph{Baseline and workloads}
AscendFHE and our baseline all use the same parameters that satisfy 128-bit security with a ring degree of $N=2^{16}$ throughout.
We evaluate the following four workloads:
\begin{enumerate}
    \item \emph{Bootstrapping (BS).} We use Cheddar's 128-bit CKKS
    preset~\cite{choi2026cheddar} with $2^{15}$ slots,
    $\Delta=2^{40}$, 47 primes in $Q$, and $dnum=4$. It refreshes a ciphertext
    from level~0 to level~13.
    \item \emph{ResNet-20}~\cite{he2016deep}. We classify one encrypted
    CIFAR-10~\cite{krizhevsky2009learning} image.
    \item \emph{BERT-base}~\cite{devlin2019bert}. We use
    THOR's configuration~\cite{moon2025thor} with sequence length~128,
    and 12 attention heads.
    \item \emph{Llama 3 8B}~\cite{meta2024llama3modelcard}. We use
    Sylph's prefill design~\cite{park2026scaling} with 128 encrypted
    tokens.
\end{enumerate}

\subsection{Overall Performance}
\label{sec:eval-end-to-end}

\textbf{End-to-end performance.}
On Ascend 910B3, AscendFHE reduces bootstrapping latency from
870 to 189\,ms, achieving a 4.61$\times$ speedup over the ported baseline.
Table~\ref{tab:eval-end-to-end} reports application speedups of
2.00$\times$, 1.71$\times$, and 1.59$\times$ for ResNet-20,
BERT-base, and Llama 3 8B, respectively.

\begin{table}[t]
\centering
\caption{End-to-end application performance relative to the Ascend baseline.}
\label{tab:eval-end-to-end}
\evaltablestyle
\begin{tabularx}{\columnwidth}{>{\raggedright\arraybackslash}Xrrr}
\toprule
\rowcolor{EvalTableHeader}
\textbf{Workload}
& \evalnumhead{Baseline (s)}
& \evalnumhead{AscendFHE (s)}
& \textbf{Speedup} \\
\midrule
ResNet-20 & 20.20 & 10.11 & \textbf{2.00$\times$} \\
BERT-base & 1072 & 626 & \textbf{1.71$\times$} \\
Llama 3 8B & 77471 & 48814 & \textbf{1.59$\times$} \\
\bottomrule
\end{tabularx}
\end{table}

\begin{figure}[t]
\centering
\includegraphics[width=\columnwidth]{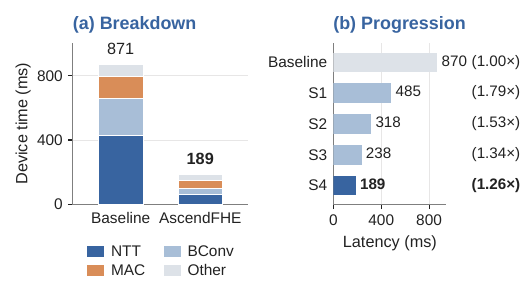}
\caption{Bootstrapping. (a) Device time. ``Other'' includes rescaling,
automorphisms, and data movement. (b) End-to-end latency and speedups
over the preceding stage.}
\label{fig:eval-bootstrap-overview}
\end{figure}

\textbf{Bootstrapping breakdown.}
Figure~\ref{fig:eval-bootstrap-overview}(a) partitions device time
into mutually exclusive intervals.
NTT/INTT and BConv occupy 75.80\% of the baseline time;
their intervals shrink from 429 and 231\,ms to 61.46 and
35.17\,ms, respectively.
Key-MAC/modular arithmetic falls from 139 to 49.34\,ms,
but its share of device time rises from 15.96\% to 26.13\%.

\textbf{Cumulative optimization gains.}
In Figure~\ref{fig:eval-bootstrap-overview}(b), \textbf{S1} moves all NTT/INTT to Cube in the Vector baseline and uses tuned Cube optimizations
and keeps all BConv calls on Vector. \textbf{S2} moves said BConv
calls from Vector to Cube and feeds their outputs directly into NTT, yielding a
1.53$\times$ speedup over S1. \textbf{S3} fuses Rescale post-processing into NTT
finalization and pipelines tasks to overlap Cube and Vector execution and gains a 1.34$\times$ speedup over S2.
Range-aware arithmetic in \textbf{S4} bring the gain over
S1 to 2.57$\times$.

\subsection{Effects of Individual Optimizations}
\label{sec:eval-optimizations}

\begin{figure}[htbp]
\centering
\includegraphics[width=\columnwidth]{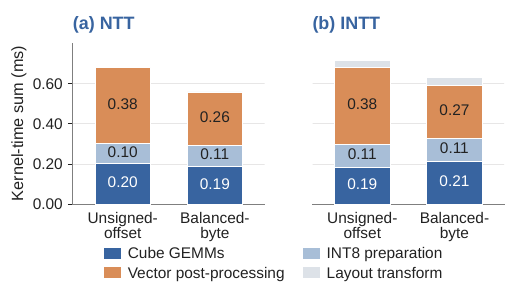}
\caption{NTT/INTT kernel-time breakdown.}
\label{fig:ntt-internal-breakdown}
\end{figure}

\textbf{Balanced-byte encoding.}
Figure~\ref{fig:ntt-internal-breakdown} sums per-stage kernel times
profiled at batch~1 with 47 RNS moduli.
INT8 preparation encodes and packs input coefficients for the first
Cube round. Layout transform accounts for the final INTT transpose.
Vector post-processing combines middle processing, input-sum
compensation, and final reconstruction/reduction.
Compared with unsigned-offset encoding, balanced-byte encoding
eliminates input-sum compensation. It accelerates middle processing
by 1.35$\times$ for both NTT and INTT, and final reconstruction/reduction
by 1.49$\times$ and 1.48$\times$, respectively.
The total kernel-time speedup is 1.22$\times$ for NTT and 1.14$\times$ for INTT.

\textbf{Consumer-driven operator handoffs.}
Table~\ref{tab:eval-operator-handoffs} reports full-batch interface latency
at batch~1 with 47 RNS moduli, including packing, device copies,
and synchronization. The baseline writes intermediate results for the next
stage to reload. The optimized paths pack INT8 inputs directly for the second
Cube round (NTT/INTT) or the following NTT (BConv). With the 12-to-47-limb
BConv on Cube in both configurations, direct handoff gives a
1.07$\times$ speedup for each variant. Full Rescale covers INTT, BConv,
NTT, and rounding. Its optimized path fuses the rounding correction
into NTT finalization. Each optimization saves one native call
and 23.50\,MiB of logical GM traffic.

\begin{table}[!htbp]
\centering
\caption{Per-call latency (ms) for operator handoffs at batch~1.}
\label{tab:eval-operator-handoffs}
\evaltablestyle
\begin{tabularx}{\columnwidth}{>{\raggedright\arraybackslash}Xrrr}
\toprule
\rowcolor{EvalTableHeader}
\textbf{Operator path}
& \textbf{Baseline}
& \textbf{Optimized}
& \textbf{Speedup} \\
\midrule
NTT round handoff
& 0.76 & 0.66 & 1.15$\times$ \\
INTT round handoff
& 0.80 & 0.70 & 1.15$\times$ \\
Canonical BConv $\to$ NTT
& 0.98 & 0.91 & 1.07$\times$ \\
Centered BConv $\to$ NTT
& 1.00 & 0.93 & 1.07$\times$ \\
Full Rescale
& 1.30 & 1.25 & 1.04$\times$ \\
\bottomrule
\end{tabularx}
\end{table}

\textbf{Cube--Vector pipelining.}
Table~\ref{tab:eval-pipeline-ablation} uses the same full-interface timing
to compare NTT/INTT schedules at batch~2 with 12 RNS moduli
and one active UB input buffer per AIV. \textbf{A} allows one
in-flight task and releases GM slots after output writes. \textbf{B}
allows two tasks under the same release policy. \textbf{C} keeps that
limit and releases slots after the last shared read for nonfinal tasks.
Together, task pipelining and early slot reuse overlap Cube and Vector
work, reducing NTT latency from 388 to 339\,$\mu$s and INTT latency
from 418 to 372\,$\mu$s. With two active UB input buffers, C takes
342\,$\mu$s for NTT and 371\,$\mu$s for INTT.

\begin{table}[!htbp]
\centering
\caption{Full-interface latency ($\mu$s) with one active UB input buffer per AIV.}
\label{tab:eval-pipeline-ablation}
\evaltablestyle
\begin{tabularx}{\columnwidth}{>{\raggedright\arraybackslash}Xrrrr}
\toprule
\rowcolor{EvalTableHeader}
\textbf{Transform} & \evalnumhead{A} & \evalnumhead{B} & \evalnumhead{C} & \textbf{Speedup} \\
\midrule
NTT & 388 & 344 & 339 & \textbf{1.15$\times$} \\
INTT & 418 & 373 & 372 & \textbf{1.12$\times$} \\
\bottomrule
\end{tabularx}
\end{table}

\subsection{Cross-Platform Comparison}
\label{sec:eval-gpu}

Table~\ref{tab:eval-gpu} compares NPU and GPU latencies at the operator and
bootstrap levels. The bootstrap ratio of 7.85$\times$ (189 and
24.07\,ms) lies within the operator range of 5.67$\times$--10.79$\times$.

\begin{table}[!htbp]
\centering
\caption{Latency comparison on H100 PCIe 80GB and
Ascend 910B3 (ms).}
\label{tab:eval-gpu}
\evaltablestyle
\begin{tabularx}{\columnwidth}{>{\raggedright\arraybackslash}Xrrr}
\toprule
\rowcolor{EvalTableHeader}
\textbf{Operation or workload}
& \evalnumhead{GPU} & \evalnumhead{NPU} & \evalnumhead{NPU/GPU} \\
\midrule
NTT, batch 1, 9 limbs
& 0.04 & 0.39 & 10.79 \\
BConv $\to$ NTT, 7 $\to$ 23 limbs
& 0.06 & 0.39 & 6.36 \\
NTT, batch 2, 19 limbs
& 0.07 & 0.44 & 6.57 \\
NTT, batch 3, 39 limbs
& 0.17 & 0.96 & 5.67 \\
Key switching, 33 limbs
& 0.54 & 3.64 & 6.68 \\
Full bootstrapping
& 24.07 & 189 & \textbf{7.85} \\
\bottomrule
\end{tabularx}
\end{table}

H100's peak INT8 matrix and INT32 addition rates are 2.56$\times$ and
5.55$\times$ those of 910B3 (Table~\ref{tab:device-compute-peaks}).
The impact of these peak rates depends on the share of transform time
spent in GEMM and in the surrounding Vector stages.
Even with balanced-byte
encoding, non-GEMM stages account for 66.76\% of summed NTT kernel
time and 66.34\% for INTT (Figure~\ref{fig:ntt-internal-breakdown}).
Encoding, reconstruction, and modular post-processing therefore
remain substantial Vector costs.

Latency also depends on how the matrix and vector engines exchange
intermediate results.
On Ascend, Cube partial sums must pass through GM before
being loaded into UB for Vector processing, incurring data
movement and synchronization costs. GPU kernels can instead
retain locally consumed intermediates in
registers~\cite{fan2025warpdrive}.

The NPU/GPU NTT latency ratio falls from 10.79$\times$ for
batch~1 with 9 prime moduli to 5.67$\times$ for batch~3 with
39 prime moduli.
Both batch size and modulus count increase across these configurations.
The trend is consistent with amortizing per-call scheduling and
synchronization costs across more transforms.

The profiles and data paths point to Vector-side processing and GM-mediated handoffs as targets for reducing NTT latency on Ascend.

\subsection{Correctness Validation}
\label{sec:eval-correctness}
\textbf{NTT/INTT.}
Our Cube-based NTT/INTT with balanced-byte encoding and
two-word reconstruction is bit-identical to the integer
reference at $N=65{,}536$ for batch~1 with 47 primes.

\textbf{Encrypted inference.}
For Llama 3 8B, the maximum absolute error  between the
decrypted and plaintext is $7.16\times10^{-5}$.

\section{Conclusion}
\label{sec:conclusion}

AscendFHE combines exact CKKS arithmetic, consumer-driven operator
handoffs, and Cube--Vector pipelining to accelerate bootstrapping and
encrypted inference on Ascend NPUs. On Ascend 910B3, it achieves a
4.61$\times$ bootstrapping speedup and 1.59$\times$--2.00$\times$
encrypted-inference speedups across ResNet-20, BERT-base, and Llama 3 8B
over our Vector-based reproduction of Cheddar.

\balance
\bibliographystyle{IEEEtran}
\bibliography{references}
\end{document}